\documentclass[conference]{IEEEtran}

\usepackage{standalone}
\usepackage[textsize=tiny]{todonotes}
\usepackage{tabularx}
\usepackage{pgfplots, pgfplotstable}
\usepackage{textcomp}
\usepackage{makecell}
\usepackage{booktabs}
\usepackage{listings}
\usepackage{lipsum}
\usepackage{multirow}
\usepackage{color}
\usepackage[T1]{fontenc}

\usepackage[version=4]{mhchem}
\usepackage{array}
\usepackage{stfloats}
\usepackage{tabularx}
\usepackage{booktabs}
\usepackage{multicol}
\usepackage{multirow}
\usepackage{lineno}
\usepackage{booktabs}
\usepackage{mathtools}
\usepackage[figuresright]{rotating}
\usepackage{amsmath}
\usepackage{amsfonts}
\usepackage{mhchem}
\usepackage{graphicx}
\usepackage{float} 
\usepackage{subfigure}
\usepackage[colorlinks,linkcolor=blue]{hyperref}
\usepackage{todonotes}
\usepackage{color} %\textcolor{red}{your text}
\usepackage{math}
\usepackage{float}
\usepackage{subfigure}

\usepackage{amssymb}
\usepackage{algorithm}
\usepackage{algorithmic}

\usepackage{bera}% optional: just to have a nice mono-spaced font
\makeatletter
\newcommand\BeraMonottfamily{%
  \def\fvm@Scale{0.85}% scales the font down
  \fontfamily{fvm}\selectfont% selects the Bera Mono font
}
\makeatother

\makeatletter
\newcommand{\linebreakand}{%
  \end{@IEEEauthorhalign}
  \hfill\mbox{}\par
  \mbox{}\hfill\begin{@IEEEauthorhalign}
}
\makeatother

\usepackage{xcolor}
\usepackage{array}
\usepackage{wasysym}
\usepackage[flushleft]{threeparttable}% (uncomment the ones you want to load)
\begin{document}
\title{ECPC-IDS:A benchmark endometrial cancer PET/CT image dataset for evaluation of semantic segmentation and detection of hypermetabolic regions}

\author{
 \IEEEauthorblockN{Dechao Tang\IEEEauthorrefmark{1},Tianmin Du\IEEEauthorrefmark{1},Deguo Ma\IEEEauthorrefmark{1},Zhiyu Ma\IEEEauthorrefmark{1},\\Hongzan Sun\IEEEauthorrefmark{2},Marcin Grzegorzek\IEEEauthorrefmark{3},Huiyan Jiang\IEEEauthorrefmark{4},Chen Li\IEEEauthorrefmark{1}}
 \IEEEauthorblockA{\IEEEauthorrefmark{1}Microscopic Image and Medical Image Analysis Group, College of Medicine and Biological Information Engineering, \\Northeastern University, Shenyang, China}
 \IEEEauthorblockA{\IEEEauthorrefmark{2}Department of Radiology, Shengjing Hospital, China Medical University, Shenyang, China}
 \IEEEauthorblockA{\IEEEauthorrefmark{3}Institute of Medical Informatics, University of Luebeck, Luebeck, Germany}
 \IEEEauthorblockA{\IEEEauthorrefmark{4}Software College, Northeastern University, Shenyang, China}
}
\maketitle

\thispagestyle{plain}
\pagestyle{plain}

\begin{abstract}

Background: Endometrial cancer is one of the most common tumors in the female reproductive system and is the third most common gynecological malignancy that causes death after ovarian and cervical cancer. Early diagnosis can significantly improve the 5-year survival rate of patients. With the development of artificial intelligence, computer-assisted diagnosis plays an increasingly important role in improving the accuracy and objectivity of diagnosis, as well as reducing the workload of doctors. However, the absence of publicly available endometrial cancer image datasets restricts the application of computer-assisted diagnostic techniques. Methods: In this paper, a publicly available \emph{Endometrial Cancer PET/CT Image Dataset for Evaluation of Semantic Segmentation and Detection of Hypermetabolic Regions} (ECPC-IDS) are published. Specifically, the segmentation section includes PET and CT images, with a total of 7159 images in multiple formats. In order to prove the effectiveness of segmentation methods on ECPC-IDS, five classical deep learning semantic segmentation methods are selected to test the image segmentation task. The object detection section also includes PET and CT images, with a total of 3579 images and XML files with annotation information. Six deep learning methods are selected for experiments on the detection task. Results: This study conduct extensive experiments using deep learning-based semantic segmentation and object detection methods to demonstrate the differences between various methods on ECPC-IDS. From a separate perspective, the minimum and maximum values of Dice on PET images are 0.471 and 0.808, respectively. The minimum and maximum values of Dice on CT images are 0.270 and 0.400, respectively. The target detection section's maximum AP values on PET and CT images are 0.993 and 0.914, respectively. Conclusion: As far as we know, this is the first publicly available dataset of endometrial cancer with a large number of multiple images, including a large amount of information required for image and target detection. ECPC-IDS can aid researchers in exploring new algorithms to enhance computer-assisted technology, benefiting both clinical doctors and patients greatly. ECPC-IDS is freely published for non-commercial at:~\url{https://figshare.com/articles/dataset/ECPC-IDS/23808258}
  
\end{abstract}

\bstctlcite{IEEEexample:BSTcontrol}

\IEEEpeerreviewmaketitle

\section{Introduction}
\label{introduction}
\subsection{Research background and motivation}
Endometrial cancer (EC) is a tumor that originates in the endometrium. In the past few decades, its incidence rate is on the rise all over the world, especially in developed countries~\cite{Morice-2016-EC}. In 2022, an estimated 84,520 new cases and 17,543 deaths are expected in China~\cite{Xia-2022-CSICA}. Endometrial cancer usually has a positive prognosis, particularly during early diagnosis, with a 5-year survival rate of over 95\% when patients are at an early stage. However, the rate significantly reduces for patients in the later stages~\cite{Choi-2018-EC}. Therefore, how to accurately diagnose in early stage is more important. \ce{^{18}F}-labeled fluoro-2-deoxyglucose (\ce{^{18}F}-FDG) positron emission tomography/computed tomography (PET/CT) is a new non-invasive diagnostic technique that can measure the extent of the tumor precisely and provide preoperative staging, which has proven to be effective~\cite{Noriega-2023-systematic}. \par

\begin{figure*}[htp!]
    \centering
    \includegraphics[width=1\linewidth]{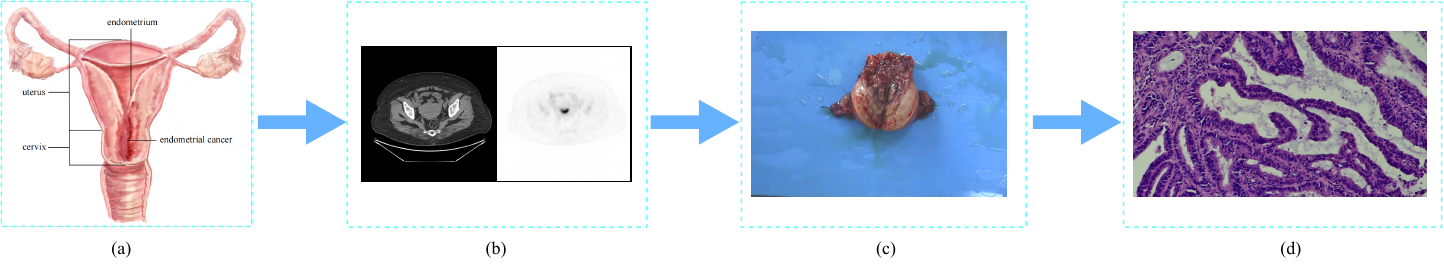}
    \caption{The treatment process of endometrial cancer, (a): Schematic diagram of endometrial cancer; (b) PET/CT imaging assisted diagnosis; (c) : Tumor tissue removed after surgery; (d) Pathological images are used for prognosis.}
    \label{fig:The treatment process of endometrial cancer}
\end{figure*}

PET technology utilizes the physical character of radioisotopes and the high metabolic biological properties of tumor regions to generate functional images. However, PET images have low spatial resolution and cannot provide anatomical information. CT images can well compensate for this weakness. The fusion of PET and CT images provides more valuable assistance to physicians~\cite{li-2020-DLFVM}. Generally, radiologists can use PET/CT sequence images to diagnose the extent and anatomy of EC. However, artificial diagnosis has shortcomings such as insufficient quantification and subjective interference. Moreover, the workload is very large, which can easily lead the doctors inefficient. Computer-assisted technology achieves the automation of tumor segmentation and target detection by extracting a large amount of image information. Through deeper mining, prediction, and analysis of massive data, it helps physicians make accurate judgments. Fig.~\ref{fig:The treatment process of endometrial cancer} shows the process of endometrial cancer from imaging diagnosis, surgery to prognosis.\par

Artificial intelligence (AI) technology can make quantitative assessments of human tissue~\cite{Sadriddinovna-2022-AIIM}. In addition, the use of AI avoids the impact of subjectivity~\cite{Fazlollahi-2022-EOAIT}. Above all, AI can work continuously and steadily, solving the problem of inefficiency caused by human fatigue~\cite{Reeder-2022-IOAIO}. However, the training of deep learning models requires a large amount of relevant data. As yet, there is still no dataset available for endometrial cancer based on PET/CT sequences. Therefore, we have produced a high-quality dataset about EC. \par 

This paper introduces a benchmark \emph{Endometrial Cancer PET/CT Image Dataset for Evaluation of Semantic Segmentation and Detection of Hypermetabolic Regions} (ECPC-IDS), which is a publicly available and highly completed PET/CT multimodality segmentation and object detection dataset for endometrial cancer, consisting of 7159 images and 3579 XML files with annotation information are included in 155 cases. Each image in this dataset calculates two different modalities and formats. In addition, a variety of major semantic segmentation and object detection deep learning methods are used to evaluate the effectiveness of the dataset. \par 

The main contributions of this paper are following:
\begin{itemize}
    \item ECPC-IDS is the first large-scale multimodality fusion dataset for endometrial cancer that can provide both semantic segmentation and object detection functions.
    \item It is proved that ECPC-IDS can be used for various classic models in deep learning, including semantic segmentation and object detection.
    \item ECPC-IDS is published as open source for non-commercial purposes. 
\end{itemize}

\subsection{Related work}
This study analyzes available research about EC, and conducts an in-depth exploration of present study results. It also points out the limitations of the current datasets on endometrial cancer.\par

Through research and extensive literature review, it is found that there is no high-quality dataset for EC with multimodality and multi format features. Now available research on endometrial cancer can be roughly grouped into two categories: other modalities such as magnetic resonance (MR) imaging for EC study and PET/CT fusion modality for other disease research. In~\cite{Chen-2020-DLFTD}, it researched myometrial invasion by EC based on MR images, including 530 patients, of whom 138 are in the test dataset. The aim was to determine the diagnostic performance of deep learning models in evaluating myometrial invasion. The accuracy of this model is higher than that of radiologists based on pathological examination (84.8\% versus 78.3\%). However, this study has limitations, including database imbalance.\par

In~\cite{Zhang-2021-DLMDC}, this study constructed a deep learning model that uses hysteroscopic images as input and can automatically classify endometrial lesions, including 1851 images from 454 patients. After image preprocessing (histogram equalization, noise addition, rotation and flips), the training set of 6478 images is input into the optimized VGG-16 model~\cite{Simonyan-2014-VGG16}, using 250 images as a test set to evaluate the performance of the model. The overall accuracy of the VGG-16 model in the classification of endometrial lesions is 80.8\%. In this task, the diagnostic performance of the VGG-16 model was slightly better than that of three gynecologists. With the help of this model, the overall accuracy of gynecologists in diagnosing endometrial lesions can be improved.\par

In~\cite{Crivellaro-2020-CPETC}, this study evaluate the application value of PET/CT in women with apparent early endometrial cancer,  including 167 patients with EC, all based on PET/CT images. The results indicate that PET/CT has high specificity for lymph node metastasis.\par
The Cancer Imaging Archive (TCIA) ~\url{https://www.cancerimagingarchive.net}, a website funded by the National Cancer Institute (NCI) cancer imaging program, is an open access database of medical images for cancer research. There are a small number of multimodality public datasets for endometrial cancer. Among them, The Cancer Genome Atlas Uterine Corpus Endometrial Carcinoma Collection (TCGA-UCEC) project provides multimodality data from 65 patients with endometrial cancer, but only 5 cases are obtained by using PET/CT fusion images. Another project, the Clinical Proteomic Tumor Analysis Consortium Uterine Corpus Endometrial Carcinoma Collection (CPTAC-UCEC), collects and provides data from 72 EC patients worldwide. Unfortunately, PET/CT images are also scarce, with only 3 cases. Moreover, these publicly available datasets have a common drawback of not providing segmentation and detection functions. The summary of related work is shown in Table~\ref{table:summary of related work}.
\begin{table*}[htp!]
    \renewcommand\arraystretch{1.8}
    \centering
    \caption{Summary of some related work.}
    \label{table:summary of related work}
\begin{tabular}{l c c c c c }
    \toprule
        \textbf{Team} & \textbf{Year} & \textbf{Reference} & \textbf{Modality} & \textbf{Aim} & \textbf{Amount} \\ 
        \midrule
        Chen et al. & 2020 & ~\cite{Chen-2020-DLFTD} & MR & detection & 530 \\ 
        Zhang et al. & 2021 & ~\cite{Zhang-2021-DLMDC} &  & classification & 454 \\
        Crivellaro et al. & 2020 & ~\cite{Crivellaro-2020-CPETC} & PET/CT & classification & 167 \\
        TCGA-UCEC & 2020 & ~\cite{TCGA-UCEC-web} & CT/MR/PET & & 72 \\
        CPTAC-UCEC & 2023 & ~\cite{CPTAC-UCEC-web} & CT/MR/PET & & 65 \\
    \bottomrule
\end{tabular}
\end{table*}

\subsection{Structure of this paper}
This section introduces the background and motivation of dataset production, and summarizes relevant research papers and datasets. Next, in Section~\ref{Data and method}, the preparation process of the ECPC-IDS is introduced, and various methods for evaluating the dataset are described in detail, including the specifics of each project in the dataset. Section~\ref{Results and discussion of semantic segmentation} and Section~\ref{Results and discussion of object detection} respectively show the experimental results and discussion of various semantic segmentation and object detection models on ECPC-IDS. Finally, a conclusion and future work plan are provided.
\section{Data and method}
\label{Data and method}
\subsection{Dataset preparation}
\subsubsection{Data source}
This study selects 155 EC patients who underwent surgery at Shengjing Hospital affiliated to China Medical University from 2013 to 2022 as the research subjects. The inclusion criteria are as follows: (1) According to the 2018 International Federation of Obstetrics and Gynecology (FIGO) classification, biopsy confirmed that cervical squamous cell carcinoma is in stage IA-IIA~\cite{Bhatla-2018-COCU}; (2) No other malignant tumors; (3) There are no preoperative radiotherapy or chemotherapy. \par

All patients receive intravenous injection of 3.7 MBq/kg \ce{^{18}F}-FDG. After a 60-minute absorption period, PET/CT (Discovery PET/CT 690; GE Healthcare) will be operated. The CT images are obtained under 120 kV tube voltage, 30-210 Ma, and 3.27mm slice thickness. PET data is collected at a rate of 1.5 minutes per bed (7-8 beds in total), with a matrix size of 192 × 192 and adopts a 3D acquisition mode. The attenuation corrected PET image is reconstructed using the iterative reconstruction algorithm with the expectation maximization of order subsets. The iterative reconstruction algorithm has two iterations, 24 subsets and 6.4 mm Gaussian filter.

\subsubsection{Rules for data preparation}
The preparation of ECPC-IDS refers to the existing methods of relevant datasets and its own characteristics, and the preparation rules are as follows: \par
Rule-\uppercase\expandafter{\romannumeral1}: The segmentation of the tumor region is performed by two experienced radiologists. Manually outline the entire tumor on each layer of endometrial cancer image using 3DSlicer software. Several researchers in biomedical at Northeastern University do not participate in delineating the edge of the lesion. Instead, researchers draw object detection box to encircle the tumor area and its surrounding structures. These bounding boxes will serve as the ground truth (GT) for the detection model. \par

Rule-\uppercase\expandafter{\romannumeral2}: It should be noted that due to the difference in pixel spacing between PET imaging and CT imaging, it is necessary to register each identical layer. After registration, the region of interest (ROI) and detection box of the lesion will be outline. \par

\begin{figure}[htp!]
    \centering
    \includegraphics[width=0.95\linewidth]{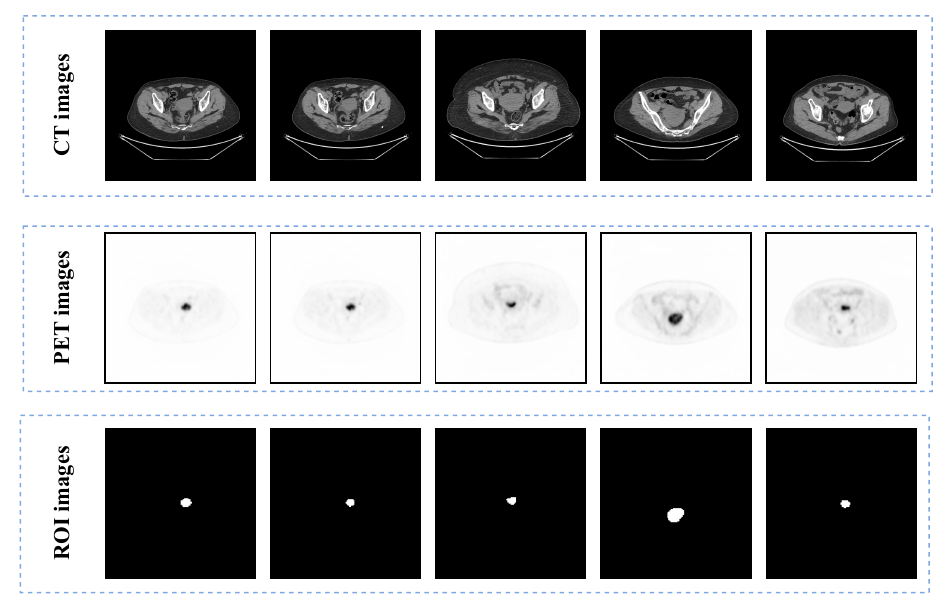}
    \caption{Randomly select CT images, PET images, and ROI images of some patients for display.}
    \label{fig: display of dataset}
\end{figure}

ECPC-IDS contains sub-sets with image segmentation and object detection functions, separately. Fig.~\ref{fig: display of dataset} is an example of some images of segmentation in ECPC-IDS. \par
\textbf{Sub-set A:} Segmentation 
\begin{itemize}
    \item [1)] ECPC-IDS is grouped into a training set (60 cases with 476 images), a validation set (61 cases with 477 images), and a testing set (34 cases with 240 images) in a 4:4:2 ratio. 
    \item[2)] Next is the preprocessing and expanding of image, including normalizing the input image, randomly rotating and scaling the input image to increase the size of the dataset.
    \item[3)] Experiment with ECPC-IDS using various classic medical image segmentation methods, and output the trained network segmented images on the test set.
    \item[4)] Finally, there is performance evaluation. This is to verify the effectiveness of segmentation, including multiple evaluation indicators.
\end{itemize} \par
\textbf{Sub-set B:} Object detection 
\begin{itemize}
    \item[1)] The training set (965 images), validation set (108 images), and test set (120 images) of ECPC-IDS are grouped in a ratio of 9:1, where the training set to validation set is also 9:1.
    \item[2)] Data preprocessing and expansion are also required, including random flipping and scrambling of the order of the real box arrangement.
    \item[3)] Experiment with various classical networks and evaluate them with various evaluation indicators.
\end{itemize}

\subsection{Dataset description}
\begin{figure}[ht]
    \centering
    \includegraphics[width=1\linewidth]{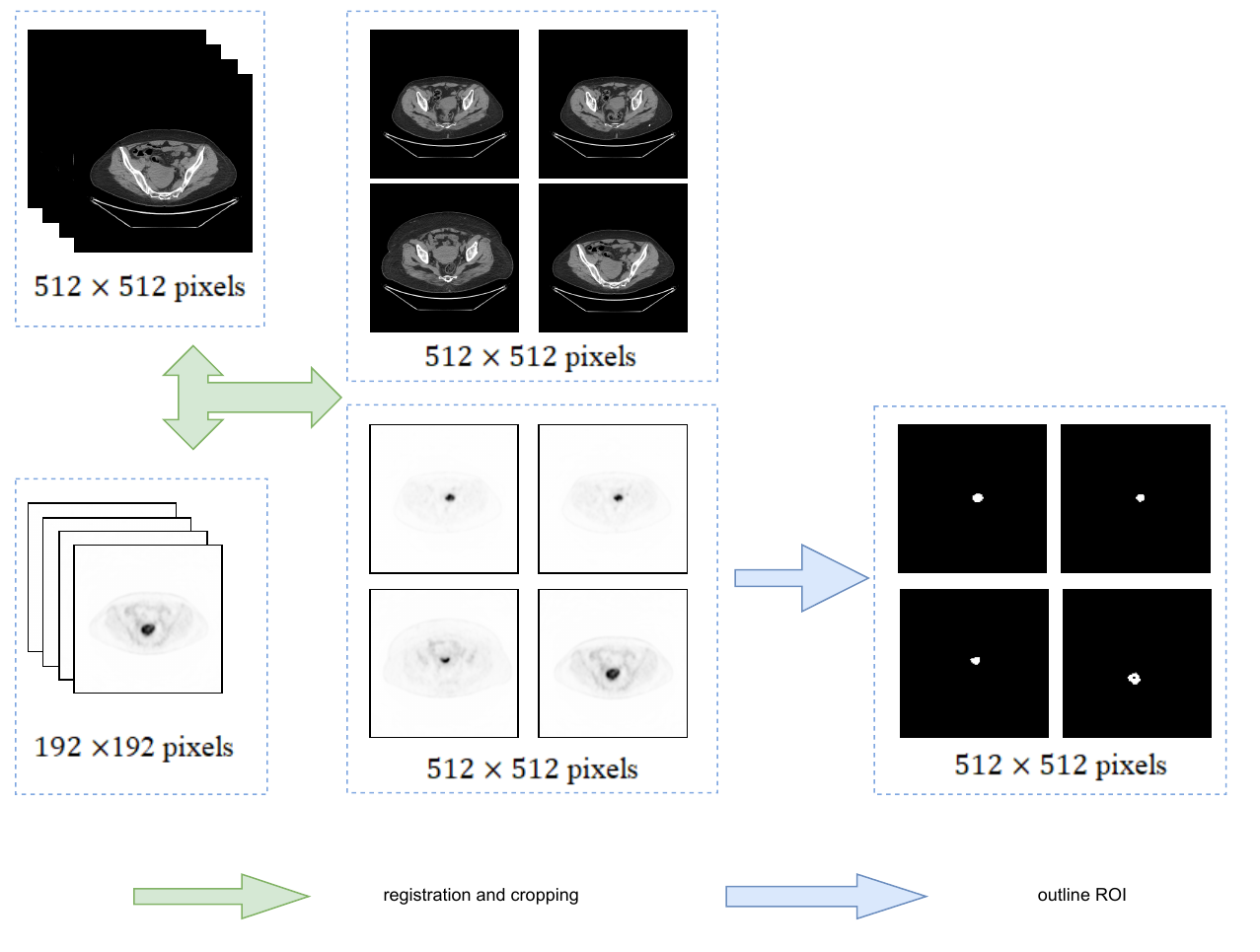}
    \caption{The workflow of ECPC-IDS.}
    \label{fig:workflow}
\end{figure}

ECPC-IDS includes PET/CT images in ``DICOM'' and ``PNG'' formats for the convenience of different research needs and the ``XML'' file with annotation information for the object detection. ECPC-IDS not only provides two formats of images, but also offers semantic segmentation and object detection functions for PET and CT fusion images of the tumor layer, respectively. Moreover, it also has the above effects on the corresponding GT images. Fig.~\ref{fig:workflow} shows the workflow of ROI production in ECPC-IDS.

%\begin{table*}[htp!]
%    \renewcommand\arraystretch{1.5}
%    \centering
%    \caption{ECPC-IDS overview.}
%   \label{table:overview of ECPC-IDS}
%\begin{tabular}{l c c c c }
%    \toprule
%        \textbf{Function} & \textbf{Format} & \textbf{CT} & \textbf{PET} & \textbf{GT} \\ 
%        \midrule
%        \multirow{2}{*}{segmentation} & DICOM & 1193 & 1193 & 1193 \\ 
%        \multirow{2}{*}{}& PNG & 1193 & 1193 & 1193 \\ 
%        \cline{2-5}
%        \multirow{2}{*}{detection} & PNG & 1193 & 1193 & 1193 \\
%        \multirow{2}{*}{}& XML & 1193 & 1193 & 1193 \\
%    \bottomrule
%\end{tabular}
%\end{table*}

\subsection{Methods of semantic segmentation and detection}
Traditional image segmentation and detection methods are no longer as effective as deep learning. Compared to traditional machine learning and computer vision methods, deep learning has significant advantages in accuracy and speed. The use of deep learning for medical image can assist medical doctors objectively confirm the size and boundaries of tumor lesions, quantitatively evaluate preoperative and postoperative effects, and greatly reduce the load of doctors. \par

\subsubsection{Semantic segmentation methods}
UNet~\cite{Ronneberger-2015-UNet} is an improvement on the FCN~\cite{Long-2015-FCN} network, which is specifically designed for medical image segmentation that  make significant progress in addressing the characteristics of small data and poor resolution, blurred edges, high noise, and complex imaging in medical images. UNet++~\cite{Zhou-2018-UNet++} is embedded with different depths of UNet, which brings better performance. In addition, it has a more flexible skip connection structure. Meanwhile, the addition of pruning operations accelerates network speed. SegNet~\cite{Badrinarayanan-2017-SegNet} uses VGG-16~\cite{Simonyan-2014-VGG16} as the main framework, removes the fully connected layer, and builds a symmetric model. The difference is that it adds indexing functionality during the pooling process, which improves the effectiveness of edge segmentation. Transformer has achieved exciting results in the field of computer vision. TransUNet~\cite{Chen-2021-TransUNet} combines UNet with Transformer~\cite{Vaswani-2017-Transformer}, and its global attention mechanism breaks through the limitations of convolutional neural networks (CNN), achieving more accurate positioning. Swin-Unet~\cite{Cao-2023-Swin-Unet} is the first pure Transformer medical image segmentation network. The tokenized image blocks are sent to the Transformer based U-shape encoder-decoder architecture through skip connections for local and global semantic feature learning, improving the spatial resolution of the feature map. \par

\subsubsection{Object detection methods}
Object detection methods can be catagorized into two groups: one stage method and two stage method. The YOLO series is currently the mainstream one stage object detection algorithm. These methods use the idea of regression, using the entire graph as input to the network, directly regressing the target border and the category of the target at multiple positions in the image. YOLOv3~\cite{Redmon-2018-YOLOv3}, YOLOv4~\cite{Bochkovskiy-2020-YOLOv4} and YOLOv5~\cite{Zhu-2021-YOLOv5} are relatively mature models. SSD~\cite{Liu-2016-SSD} is another one stage model type, unlike YOLO, which uses a fully connected layer at the end, SSD directly uses convolution to extract detection results from different feature maps. Meanwhile, SSD draws on the concept of anchors in Faster R-CNN~\cite{Ren-2015-FasterRCNN}. SSD not only maintains the fast speed feature of YOLO, but also ensures accuracy like Faster RCNN. The emergence of RetinaNet~\cite{Lin-2017-RetinaNet} make one stage networks surpass two stage networks for the first time, with its most important contribution being focal loss. The aggregator of the two stage type is Faster R-CNN. Faster R-CNN is an improvement on R-CNN~\cite{Girshick-2014-RCNN} and Fast R-CNN~\cite{Girshick-2015-FastRCNN}, which significantly improves the accuracy and speed of detection after introducing the region proposal networks (RPN) framework. Although its accuracy is considerable, its speed cannot meet the real-time requirements.

\subsection{Evaluation metrics}
A correct objective indicator is required to measure the effectiveness of various models on the dataset. In medical images segmentation, the golden standard (ground truth or GT) is usually doctor's hand-drawn annotated images. The algorithm's segmentation result is the predicted image (Rseg or SEG). ECPC-IDS refers to eight commonly used indicators for evaluating medical image segmentation and object detection. \par

\textbf{Dice index:} It is the most commonly used to evaluate similarity, such as the degree of similarity or overlap between two samples. Its value range is $[0,1]$. The closer the value is to 0, the worse the segmentation effect. On the contrary, the closer it is to 1, the better the segmentation effect. Given two sets A and B, the metric is defined as Eq.(~\ref{eq:Dice} ) \par

\begin{equation}
  \mathrm{Dice(A,B)}=2\frac{\mathrm{|A\cap B|}}{\mathrm{|A| + |B|}}
  \label{eq:Dice}
\end{equation}

\textbf{Jaccard index :} It is similar to the dice coefficient. Given two sets A and B, the calculation formula is Eq.(~\ref{eq:Jaccard} ) \par
\begin{equation}
    \mathrm{Jaccard(A,B)}=\frac{\mathrm{|A\cap B|}}{\mathrm{|A\cup B|}}
    \label{eq:Jaccard}
\end{equation}

\textbf{Hausdorff distance:} It describes a measure of the similarity between two sets of points, that is, the distance between the two boundaries of the GT and the segmentation results predicted by the model. namely, the sensitivity to the segmentation boundary, its definition is as shown in Eq.(~\ref{eq:Hd} )
\begin{equation}
  \mathrm{Hd}=\mathrm{max}\bigg(\underset{i\in \mathrm{seg}}{\mathrm{max}}\bigg(\underset{j\in \mathrm{gt}}{\mathrm{min}}\big(d(i,j)\big)\bigg),\underset{j\in \mathrm{gt}}{\mathrm{max}}\bigg(\underset{i\in \mathrm{seg}}{\mathrm{min}}\big(d(i,j)\big)\bigg) \bigg)
  \label{eq:Hd}
\end{equation}

where, \emph{i} and \emph{j} represent points from different sets, respectively, \emph{d} is the distance between \emph{i} and \emph{j}, \emph{seg} means the predicted result, and \emph{gt} is the gold standard.\par
\textbf{Precision}, \textbf{Recall}, and \textbf{F1-score} are three other major indicators. Their definitions are as Eq.(~\ref{eq:Precision} ), Eq.(~\ref{eq:Recall} ) and Eq.(~\ref{eq:F1} ) \par

\begin{equation}
    \mathrm{Precision}=\frac{\mathrm{TP}}{\mathrm{TP + FP}}
    \label{eq:Precision}
\end{equation}

\begin{equation}
    \mathrm{Recall}=\frac{\mathrm{TP}}{\mathrm{TP + FN}}
    \label{eq:Recall}
\end{equation}

\begin{equation}
    \mathrm{F1}=\frac{2 \mathrm{TP}}{2 \mathrm{TP + FP + FN}}
    \label{eq:F1}
\end{equation}

The definitions of TP, FP, TN, and FN are shown in Table~\ref{table:confusion matrix}
\begin{table}[h!]
    \renewcommand\arraystretch{1.5}
    \centering
    \caption{Confusion matrix.}
    \label{table:confusion matrix}
\begin{tabular}{|c|c|c|}
    \hline
    \multirow{2}{*}{Ground truth} &\multicolumn{2}{|c|}{Predict mask} \\
    \cline{2-3}
    & Positive & Negative \\
    \cline{1-3}
    Positive & TP & TN \\
    \cline{1-3}
    Negative & FP & FN \\
    \hline
\end{tabular}
\end{table} \par
The conformity coefficient (\textbf{Confm}) is used to calculate the ratio between the number of correctly segmented pixels and the number of wrong segmented pixels, in order to measure the consistency between the segmentation results and ground truth. Its definition is as Eq.(~\ref{eq:Confm1} ) and Eq.(~\ref{eq:Confm2} ) \par
\begin{equation}
 \mathrm{Confm}=(1- \frac{\mathrm{\theta_ {AE} }}{\theta_{\mathrm{TP}} }) ,\theta_{\mathrm{TP}} > 0 
 \label{eq:Confm1}
\end{equation}

\begin{equation}
\mathrm{Confm}=\mathrm{Failure},\theta_{\mathrm{TP}}=0
\label{eq:Confm2}
\end{equation}

where $\theta_{\mathrm{AE}}$=$\theta_{\mathrm{FP}} $+ $\theta_{\mathrm{FN}}$, it means all errors in segmentation. $\theta_{\mathrm{TP}}$ is the correct sum in pixel classification. If $\theta_{TP}=0$, then Confm will be infinitely close to negative infinity, which means that the segmentation result is insufficient and can be considered a failure. \par
Average precision (\textbf{AP}) actually refers to the area below the curve drawn using a combination of different Precision and Recall points, and the \textbf{mAP} is the mean value.
\section{Results and discussion of semantic segmentation}
\label{Results and discussion of semantic segmentation}
\subsection{Results of semantic segmentation model}
In this part, classic segmentation models based on deep learning are used to experiment on the sub-set A dataset of ECPC-IDS. In a series of comparative experiments, each model use a learning rate of 0.001, set the batch size to 4, and conduct experiments for 100 and 200 epochs to observe the effectiveness of ECPC-IDS on different models. The results of the comparative experiment are shown in Table~\ref{table:results of PET images in subsetA} and Table ~\ref{table:results of CT images in subsetA}. \par
For semantic segmentation, the experiments are conducted using a local workstation, with a GPU of 16 GB NVIDIA RTX 2080 and 32GB RAM, running on the Windows 10 oprating system. As for the software part, Python 3.7 is used for programming and Python version 1.7.1 is used as the framework. \par

\begin{table*}[htp!]
    \renewcommand\arraystretch{1.9}
    \centering
    \caption{The results of using different deep learning based semantic segmentation models in the PET modality of the sub-set A dataset of ECPC-IDS. The bold text in the table represents the best results under the same evaluation criteria.}
    \label{table:results of PET images in subsetA}
\resizebox{\linewidth}{!}{\begin{tabular}{ m{0.1\textwidth}<{\raggedright} m{0.08\textwidth}<{\centering} p{0.1\textwidth}<{\centering} m{0.1\textwidth}<{\centering} m{0.15\textwidth}<{\raggedright} m{0.08\textwidth}<{\centering} c c c c c c }
\toprule
        \textbf{Sub-dataset} & \textbf{Modality} & \textbf{Model} & \textbf{Quantity of epoch} & \textbf{Training time} & \textbf{Testing time/per image} & \textbf{Dice} & \textbf{Jaccard} & \textbf{Hd} & \textbf{Confm} & \textbf{Precision} & \textbf{Recall} \\ 
\midrule
        \multirow{10}{*}{Sub-set A} & \multirow{10}{*}{PET} & UNet & 100 & 2847s ($\approx$47min) & 0.554s & 0.697 & 0.535 & 8.602 & 0.132 & 0.963 & 0.547 \\ 
        \multirow{10}{*}{} & \multirow{10}{*}{} &  & 200 & 5640s($\approx$1h34min) & 0.558s & 0.634 & 0.467 & 9.900 & -0.143 & \textbf{0.966} & 0.475 \\  [1.5 ex]
        \multirow{10}{*}{} & \multirow{10}{*}{} & UNet++ & 100 & 3021s($\approx$50min) & 0.583s & 0.733 & 0.578 & 8.602 & 0.270 & 0.893 & 0.621 \\
        \multirow{10}{*}{}& \multirow{10}{*}{} &  & 200 & 6065s($\approx$1h51min) & 0.583s & 0.739 & 0.586 & 55 & 0.293 & 0.713 & 0.767 \\ [1.5 ex]
        \multirow{10}{*}{} & \multirow{10}{*}{} & SegNet & 100 & 2417s($\approx$40min) & 0.483s & 0.471 & 0.308 & 33.38 & -1.248 & 0.382 & 0.614 \\
        \multirow{10}{*}{} & \multirow{10}{*}{} &  & 200 & 4826s($\approx$1h20min) & 0.475s & 0.565 & 0.394 & 49 & -0.542 & 0.597 & 0.536 \\ [1.5 ex]
        \multirow{10}{*}{} & \multirow{10}{*}{} & TransUNet & 100 & 3012s($\approx$50min) & 0.442s & \textbf{0.808} & \textbf{0.679} & 5.100 & 0.526 & 0.738 & \textbf{0.894} \\
        \multirow{10}{*}{} & \multirow{10}{*}{} &  & 200 & 6019s($\approx$1h40min) & 0.446s & 0.743 & 0.591 & 8.602 & 0.309 & 0.810 & 0.686 \\ [1.5 ex]
        \multirow{10}{*}{} & \multirow{10}{*}{} & SwinUnet & 100 & 3670s($\approx$1h1min) & 0.583s & 0.712 & 0.553 & 8.602 & 0.193 & 0.917 & 0.583 \\
        \multirow{10}{*}{} & \multirow{10}{*}{} &  & 200 & 7367s($\approx$2h2min) & 0.600s & 0.719 & 0.561 & 8.602 & 0.219 & 0.918 & 0.591 \\
\bottomrule
\end{tabular}
}
\end{table*}

\begin{table*}[htp!]
    \renewcommand\arraystretch{1.9}
    \centering
    \caption{The results of using different deep learning based semantic segmentation models in the CT modality of the sub-set A dataset of ECPC-IDS. The bold text in the table represents the best results under the same evaluation criteria.}
    \label{table:results of CT images in subsetA}
\resizebox{\linewidth}{!}{\begin{tabular}{ m{0.1\textwidth}<{\raggedright} m{0.08\textwidth}<{\centering} p{0.1\textwidth}<{\centering} m{0.1\textwidth}<{\centering} m{0.15\textwidth}<{\raggedright} m{0.08\textwidth}<{\centering} c c c c c c }
\toprule
        \textbf{Sub-dataset} & \textbf{Modality} & \textbf{Model} & \textbf{Quantity of epoch} & \textbf{Training time} & \textbf{Testing time/per image} & \textbf{Dice} & \textbf{Jaccard} & \textbf{Hd} & \textbf{Confm} & \textbf{Precision} & \textbf{Recall} \\ 
\midrule
        \multirow{10}{*}{Sub-set A} & \multirow{10}{*}{CT} & UNet & 100 & 2845s($\approx$47min) & 0.529s & 0.457 & 0.335 & 15.742 & -3.881 & 0.605 & 0.462 \\ 
        \multirow{10}{*}{} & \multirow{10}{*}{} &  & 200 & 5657s($\approx$1h34min) & 0.533s & 0.510 & 0.389 & 18.630 & -4.006 & 0.581 & 0.518 \\  [1.5 ex]
        \multirow{10}{*}{} & \multirow{10}{*}{} & UNet++ & 100 & 3026s($\approx$50min) & 0.567s & 0.503 & 0.375 & 16.07 & -4.405 & 0.657 & 0.463 \\
        \multirow{10}{*}{}& \multirow{10}{*}{} &  & 200 & 6053s($\approx$1h41min) & 0.558s & 0.472 & 0.344 & 18.63 & -2.773 & \textbf{0.701} & 0.414 \\ [1.5 ex]
        \multirow{10}{*}{} & \multirow{10}{*}{} & SegNet & 100 & 2429s($\approx$40min) & 0.458s & 0.362 & 0.253 & 19.21 & -2.774 & 0.639 & 0.319 \\
        \multirow{10}{*}{} & \multirow{10}{*}{} &  & 200 & 4704s($\approx$1h18min) & 0.458s & 0.417 & 0.305 & 21.75 & -2.557 & 0.616 & 0.401 \\ [1.5 ex]
        \multirow{10}{*}{} & \multirow{10}{*}{} & TransUNet & 100 & 3046s($\approx$51min) & 0.442s & \textbf{0.522} & \textbf{0.419} & 20.38 & -4.045 & 0.647 & \textbf{0.551} \\
        \multirow{10}{*}{} & \multirow{10}{*}{} &  & 200 & 6090s($\approx$1h42min) & 0.438s & 0.437 & 0.312 & 19.84 & -2.005 & 0.633 & 0.393 \\ [1.5 ex]
        \multirow{10}{*}{} & \multirow{10}{*}{} & SwinUnet & 100 & 3774s($\approx$1h3min) & 0.613s & 0.330 & 0.231 & 18.23 & -2.522 & 0.628 & 0.266 \\
        \multirow{10}{*}{} & \multirow{10}{*}{} &  & 200 & 7314s($\approx$2h2min) & 0.613s & 0.419 & 0.301 & 17.98 & -2.246 & 0.641 & 0.367 \\
\bottomrule
\end{tabular}
}
\end{table*}

\subsection{Discussion of semantic segmentation model}
\begin{figure*}[htp!]
    \centering
    \includegraphics[width=1\linewidth]{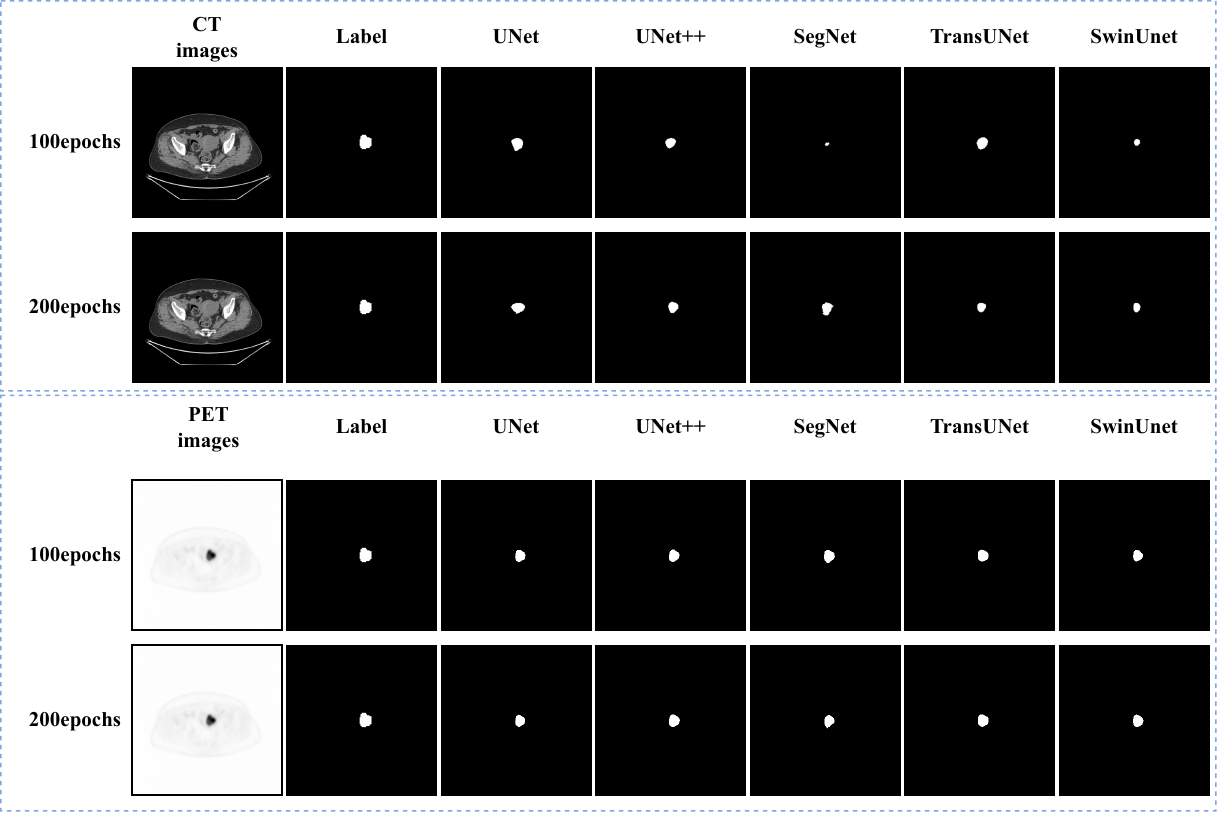}
    \caption{Comparison of predicted images using different segmentation methods on CT and PET images of the same patient.}
    \label{fig:example of segmentation}
\end{figure*}

The semantic segmentation methods based on deep learning are used to analyze the data of subset A in ECPC-IDS, and the results in Table~\ref{table:results of PET images in subsetA} and Table~\ref{table:results of CT images in subsetA} are obtained. Fig.~\ref{fig:example of segmentation} is an example of a patient's CT and PET images predicted using different segmentation methods after training. Based on the four metrics of Dice, Jaccard, Precision, and Recall, TransUNet model performs relatively well on the dataset. Convolutional neural networks represented by UNet and UNet++ often have limitations in establishing remote dependency relationships. Although the Transformer used for sequence to sequence prediction has a global attention mechanism, it lacks underlying details, resulting in limited localization capabilities. TransUNet combines the advantages of both UNet and Transformer. The cost is that the model parameters are larger and the training time is increased, respectively. SwinUnet has a preferable generalization ability from a data perspective, benefiting from its ability to remove its CNN encoding and replace its vision transformer (ViT)~\cite{Dosovitskiy-2020-ViT} with Swin Transformer compared to TransUNet. As a result, it delivers better performance with a global attention mechanism. In addition, the performance of the same model on PET and CT images differs greatly, the segmentation effect on PET is significantly better than that on CT. This is mainly due to the image structure, which means that endometrial cancer, as a soft tissue tumor, is not significantly different from surrounding normal tissues on CT. \par

\section{Results and discussion of object detection}
\label{Results and discussion of object detection}
\subsection{Results of object detection model}
In this section, classic and novel object detection models based on deep learning methods are used for relevant experiments on the sub-set B dataset of ECPC-IDS. In the experiment, the initial learning rate of each model is set to 0.01 and the stochastic gradient descent (SGD) optimization method is used. Meanwhile, the batch size is set to 8, and conduct 300 epochs of experiments to observe the performance of the dataset on different models. The results of the comparative experiment are shown in Table ~\ref{table:results of PET images in subsetB} and Table ~\ref{table:results of CT images in subsetB}. \par
The experiment in this part is the same as the segmentation experiment, using a local workstation with a GPU of 16GB NVIDIA RTX 2080, RAM of 32GB, operating system of Windows 10, programming in Python 3.7, and using Python 1.7.1 version as the framework. 

\begin{table*}[htp!]
    \renewcommand\arraystretch{1.6}
    \centering
    \caption{The results of using different deep learning based object detection models in the PET modality of the sub-set B dataset of ECPC-IDS. The bold text in the table represents the best results under the same evaluation criteria}
    \label{table:results of PET images in subsetB}
\resizebox{\linewidth}{!}{\begin{tabular}{l  m{0.08\textwidth}<{\centering} c m{0.1\textwidth}<{\centering} m{0.15\textwidth}<{\raggedright}  m{0.1\textwidth}<{\centering} c c c c }
    \toprule
        \textbf{Sub-dataset} & \textbf{Modality} & \textbf{Model} & \textbf{Quantity of epoch} & \textbf{Training time} & \textbf{Testing time/per image} & \textbf{Precision} & \textbf{Recall} & \textbf{mAP} & \textbf{F1-score}  \\ 
        \midrule
         \multirow{6}{*}{Sub-set B} & \multirow{6}{*}{PET} & YOLOv3 & 300 & 23591s($\approx$6h33min) & 0.104s & \textbf{0.983} & 0.975 & 0.988 & 0.98 \\ [1 ex]
        \multirow{6}{*}{} & \multirow{6}{*}{} & YOLOv4 & 300 & 57610s($\approx$16h) & 0.105s & 0.975 & 0.958 & 0.967 & 0.97 \\ [1 ex]
        \multirow{6}{*}{} & \multirow{6}{*}{} & YOLOv5 & 300 & 7321s($\approx$2h2min) & 0.087s & 0.96 & \textbf{0.992} & \textbf{0.992} & \textbf{0.98} \\ [1 ex]
        \multirow{6}{*}{} & \multirow{6}{*}{} & RetinaNet & 300 & 26531s($\approx$7h22min) & 0.111s & 0.975 & 0.975 & \textbf{0.993} & 0.97  \\ [1 ex]
        \multirow{6}{*}{} & \multirow{6}{*}{} & SSD & 300 & 7263s($\approx$2h1min) & 0.081s & 0.976 & 0.683 & 0.981 & 0.80 \\ [1 ex]
        \multirow{6}{*}{} & \multirow{6}{*}{} & Faster R-CNN & 300 & 37808s($\approx$10h30min) & 0.157s & 0.541 & 0.983 & 0.954 & 0.70 \\ 
    \bottomrule
\end{tabular}
}
\end{table*}

\begin{table*}[htp!]
    \renewcommand\arraystretch{1.6}
    \centering
    \caption{The results of using different deep learning based object detection models in the CT modality of the sub-set B dataset of ECPC-IDS. The bold text in the table represents the best results under the same evaluation criteria}
    \label{table:results of CT images in subsetB}
\resizebox{\linewidth}{!}{\begin{tabular}{l  m{0.08\textwidth}<{\centering} c m{0.1\textwidth}<{\centering}  m{0.15\textwidth}<{\raggedright} m{0.1\textwidth}<{\centering} c c c c }
    \toprule
        \textbf{Sub-dataset} & \textbf{Modality} & \textbf{Model} & \textbf{Quantity of epoch} & \textbf{Training time} & \textbf{Testing time/per image} & \textbf{Precision} & \textbf{Recall} & \textbf{mAP} & \textbf{F1-score}  \\ 
        \midrule
         \multirow{6}{*}{Sub-set B} & \multirow{6}{*}{CT} & YOLOv3 & 300 & 23402s($\approx$6h30min) & 0.106s & 0.782 & 0.358 & 0.615 & 0.49 \\ [1 ex]
        \multirow{6}{*}{} & \multirow{6}{*}{} & YOLOv4 & 300 & 57608s($\approx$16h) & 0.107s & 0.636 & 0.058 & 0.491 & 0.11 \\ [1 ex]
        \multirow{6}{*}{} & \multirow{6}{*}{} & YOLOv5 & 300 & 16205s($\approx$4h30min) & 0.095s & 0.83 & 0.692 & 0.743 & 0.75 \\ [1 ex]
        \multirow{6}{*}{} & \multirow{6}{*}{} & RetinaNet & 300 & 26640s($\approx$7h24min) & 0.114s & 0.915 & 0.892 & 0.911 & \textbf{0.90} \\ [1 ex]
        \multirow{6}{*}{} & \multirow{6}{*}{} & SSD & 300 & 7259s($\approx$2h1min) & 0.088s & \textbf{0.939} & 0.767 & \textbf{0.914} & 0.84 \\ [1 ex]
        \multirow{6}{*}{} & \multirow{6}{*}{} & Faster R-CNN & 300 & 39612s($\approx$11h) & 0.155s & 0.511 & \textbf{0.942} & 0.875 & 0.66 \\ 
    \bottomrule
\end{tabular}
}
\end{table*}

\subsection{Discussion of object detection model}
\begin{figure*}[htp!]
    \centering
    \includegraphics[width=1\linewidth]{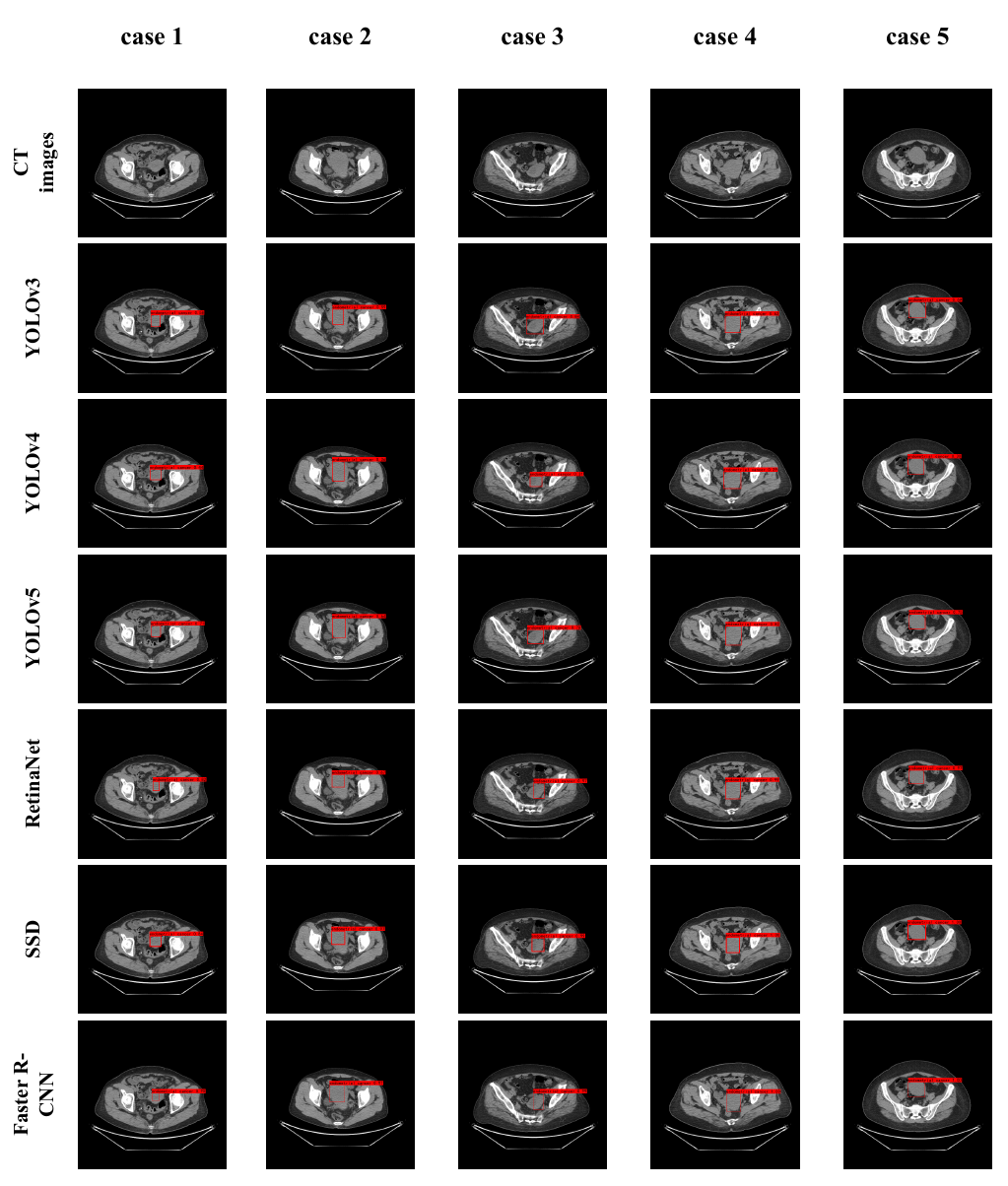}
    \caption{Comparison of prediction boxes on CT images of five patients using different object detection methods.}
    \label{fig:CT_detection}
\end{figure*}

\begin{figure*}[htp!]
    \centering
    \includegraphics[width=1\linewidth]{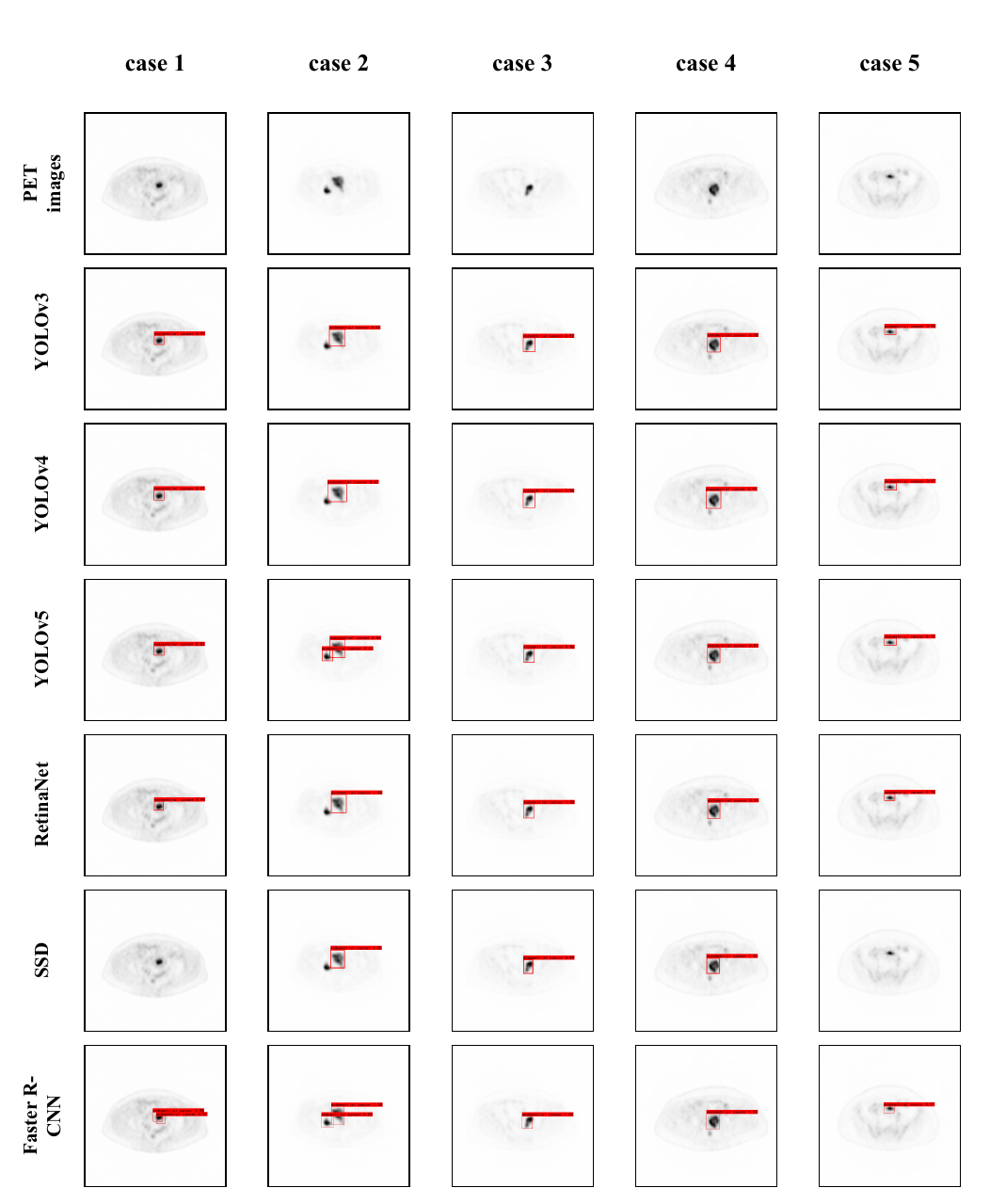}
    \caption{Comparison of prediction boxes on PET images of five patients using different object detection methods.}
    \label{fig:PET_detection}
\end{figure*}

Based on the four indicators in Tables~\ref{table:results of PET images in subsetB} and ~\ref{table:results of CT images in subsetB}, as well as the randomly selected CT and PET images of five patients shown in Fig.~\ref{fig:CT_detection} and Fig.~\ref{fig:PET_detection}, examples of predicted results using different detection models are presented. Although there are omissions, the same object detection method performs better on PET images than on CT images. In PET modality, YOLOv5 demonstrates outstanding performance, having further improved detection accuracy, especially in its ability to detect small objects, while maintaining its speed advantage. Unfortunately, Faster R-CNN does not gain an advantage in precision. In genera, the YOLO series of methods have great advantages in detecting small target objects with simple backgrounds, both in terms of accuracy and speed. In CT modality, for target objects with complex settings, the YOLO series of methods appear average. SSD is better than other one stage methods. The reason for this result is to introduce the pyramid feature hierarchy module~\cite{Lin-2017-FPN}. The introduction of this module enables SSD to perform well in detecting small targets, but the disadvantage is that the recall rate is relatively low.

\section{Conclusion and futures works}
\label{conclusion and futures works}
This paper introduces ECPC-IDS, a multimodality image dataset publicly available for endometrial cancer. ECPC-IDS contains two sub-datasets: sub-set A,which has a semantic segmentation function, and sub-set B, which has an object detection function. Each sub-dataset has two folders for PET and CT images, with cropped images that have been rearranged and registered. The dataset contains a total of 7159 images in two formats, PNG and DICOM formats, and 3579 XML files with annotation information. This paper uses different methods for segmentation experiments and uses the segmentation results for evaluation index analysis, as well as representative object detection methods for experiments. For the segmentation experiment, five classic methods and six evaluation indicators are used. To explore the optimization stages of different models, experiments are conducted on 100 and 200 epochs, respectively. In PET modality, the highest and lowest dice ratios are 0.808 and 0.471, respectively. The highest and lowest values of precision are 0.966 and 0.382, while the highest and lowest values of recall are 0.894 and 0.475, respectively. In the corresponding CT modality, the highest and lowest dice ratios are 0.400 and 0.270. The highest values of Precision and Recall are 0.264 and 0.822, while the lowest values are 0.171 and 0.470, respectively. The segmentation experiment using ECPC-IDS shows that this dataset can effectively perform segmentation tasks using each segmentation method. In addition, there are significant differences between segmentation evaluation indicators. Therefore, ECPC-IDS is practical and effective in performing image segmentation tasks. \par

For the experiment of object detection, six methods are used. In the PET modality, the highest and lowest mAP are 0.993 and 0.954. The highest values of Precision and Recall are 0.983 and 0.992, while the lowest values are 0.541 and 0.683, respectively. In the CT modality, the highest and lowest mAP are 0.914 and 0.491. The highest values of Precision and Recall are 0.939 and 0.942, while the lowest values are 0.511 and 0.058, respectively. Various indicators can indicate that ECPC-IDS has significant differentiation and effectiveness. The object detection experiment using ECPC-IDS shows that this dataset can effectively perform detection tasks in various detection methods. Furthermore, there are significant differences between the evaluation indicators obtained. Thus, ECPC-IDS is effective in performing image object detection tasks.\par

The production of ECPC-IDS means that more image segmentation and object detection methods can be applied to that dataset. We will also aim to improve the image segmentation and object detection methods on ECPC-IDS, compare and analyze image segmentation and object detection methods, and obtain better practical methods to contribute to medical progress.
\section*{Declaration of competing interes}
The authors declare that they have no conflict of interest in this paper.
\section*{Acknowledgements}
This work is supported by National Natural Science Foundation of China (No. 82220108007). We thank Miss Zixian Li and Mr. Guoxian Li for their important discussion. We also thank B.A. Yingying Hou from Foreign Studies College in Northeastern University, China, for her professional English proofreading in this paper.
\section*{Data availability statement}
Ethics is proved by China Medical Universiy, China: No. 2022PS433K. The datasets presented
in this study can be found in online repositories. The names of the repository and accession number can be found below:~\url{https://figshare.com/articles/dataset/ECPC-IDS/23808258}

\bibliographystyle{IEEEtran}
\bibliography{Tang}

\begin{thebibliography}{10}
\providecommand{\url}[1]{#1}
\csname url@samestyle\endcsname
\providecommand{\newblock}{\relax}
\providecommand{\bibinfo}[2]{#2}
\providecommand{\BIBentrySTDinterwordspacing}{\spaceskip=0pt\relax}
\providecommand{\BIBentryALTinterwordstretchfactor}{4}
\providecommand{\BIBentryALTinterwordspacing}{\spaceskip=\fontdimen2\font plus
\BIBentryALTinterwordstretchfactor\fontdimen3\font minus
  \fontdimen4\font\relax}
\providecommand{\BIBforeignlanguage}[2]{{%
\expandafter\ifx\csname l@#1\endcsname\relax
\typeout{** WARNING: IEEEtran.bst: No hyphenation pattern has been}%
\typeout{** loaded for the language `#1'. Using the pattern for}%
\typeout{** the default language instead.}%
\else
\language=\csname l@#1\endcsname
\fi
#2}}
\providecommand{\BIBdecl}{\relax}
\BIBdecl

\bibitem{Morice-2016-EC}
P.~Morice, A.~Leary, C.~Creutzberg, N.~Abu-Rustum, and E.~Darai, ``Endometrial
  cancer,'' \emph{The Lancet}, vol. 387, no. 10023, pp. 1094--1108, 2016.

\bibitem{Xia-2022-CSICA}
C.~Xia, X.~Dong, H.~Li, M.~Cao, D.~Sun, S.~He, F.~Yang, X.~Yan, S.~Zhang,
  N.~Li, W.~Chen, and J.~Ni, ``Cancer statistics in china and united states,
  2022: profiles, trends, and determinants,'' \emph{Chinese Medical Journal},
  vol. 135, no.~05, pp. 584--590, 2022.

\bibitem{Choi-2018-EC}
S.~Choi and I.-C.~J. Hsu, ``Endometrial cancer,'' \emph{Handbook of
  Evidence-Based Radiation Oncology}, pp. 653--677, 2018.

\bibitem{Noriega-2023-systematic}
E.~Noriega-{\'A}lvarez, A.~M.~G. Vicente, G.~A.~J. Londo{\~n}o, W.~R.~M. Bravo,
  B.~G. Garc{\'\i}a, and {\'A}.~M.~S. Castrej{\'o}n, ``A systematic review
  about the role of preoperative 18f-fdg pet/ct for prognosis and risk
  stratification in patients with endometrial cancer,'' \emph{Revista
  Espa{\~n}ola de Medicina Nuclear e Imagen Molecular (English Edition)},
  vol.~42, no.~1, pp. 24--32, 2023.

\bibitem{li-2020-DLFVM}
L.~Li, X.~Zhao, W.~Lu, and S.~Tan, ``Deep learning for variational
  multimodality tumor segmentation in pet/ct,'' \emph{Neurocomputing}, vol.
  392, pp. 277--295, 2020.

\bibitem{Sadriddinovna-2022-AIIM}
N.~F. Sadriddinovna, R.~F. Salimovna, and A.~B.~A. Ugli, ``Artificial
  intelligence in medicine,'' \emph{Web of Scientist: International Scientific
  Research Journal}, vol.~3, no.~5, pp. 23--27, 2022.

\bibitem{Fazlollahi-2022-EOAIT}
A.~M. Fazlollahi, M.~Bakhaidar, A.~Alsayegh, R.~Yilmaz, A.~Winkler-Schwartz,
  N.~Mirchi, I.~Langleben, N.~Ledwos, A.~J. Sabbagh, K.~Bajunaid \emph{et~al.},
  ``Effect of artificial intelligence tutoring vs expert instruction on
  learning simulated surgical skills among medical students: a randomized
  clinical trial,'' \emph{JAMA Network Open}, vol.~5, no.~2, pp.
  e2\,149\,008--e2\,149\,008, 2022.

\bibitem{Reeder-2022-IOAIO}
K.~Reeder and H.~Lee, ``Impact of artificial intelligence on us medical
  students' choice of radiology,'' \emph{Clinical imaging}, vol.~81, pp.
  67--71, 2022.

\bibitem{Chen-2020-DLFTD}
X.~Chen, Y.~Wang, M.~Shen, B.~Yang, Q.~Zhou, Y.~Yi, W.~Liu, G.~Zhang, G.~Yang,
  and H.~Zhang, ``Deep learning for the determination of myometrial invasion
  depth and automatic lesion identification in endometrial cancer mr imaging: a
  preliminary study in a single institution,'' \emph{European radiology},
  vol.~30, pp. 4985--4994, 2020.

\bibitem{Zhang-2021-DLMDC}
Y.~Zhang, Z.~Wang, J.~Zhang, C.~Wang, Y.~Wang, H.~Chen, L.~Shan, J.~Huo, J.~Gu,
  and X.~Ma, ``Deep learning model for classifying endometrial lesions,''
  \emph{Journal of Translational Medicine}, vol.~19, pp. 1--13, 2021.

\bibitem{Simonyan-2014-VGG16}
K.~Simonyan and A.~Zisserman, ``Very deep convolutional networks for
  large-scale image recognition,'' \emph{arXiv preprint arXiv:1409.1556}, 2014.

\bibitem{Crivellaro-2020-CPETC}
C.~Crivellaro, C.~Landoni, F.~Elisei, A.~Buda, M.~Bonacina, T.~Grassi,
  L.~Monaco, D.~Giuliani, I.~Gotuzzo, S.~Magni \emph{et~al.}, ``Combining
  positron emission tomography/computed tomography, radiomics, and sentinel
  lymph node mapping for nodal staging of endometrial cancer patients,''
  \emph{International Journal of Gynecologic Cancer}, vol.~30, no.~3, 2020.

\bibitem{TCGA-UCEC-web}
M.~Erickson and Lippmann, ``The cancer genome atlas uterine corpus endometrial
  carcinoma collection,''
  \url{https://wiki.cancerimagingarchive.net/pages/viewpage.action?pageId=19039602}.

\bibitem{CPTAC-UCEC-web}
P.~Wilson, ``The clinical proteomic tumor analysis consortium uterine corpus
  endometrial carcinoma collection,''
  \url{https://wiki.cancerimagingarchive.net/pages/viewpage.action?pageId=33948263}.

\bibitem{Bhatla-2018-COCU}
N.~Bhatla, D.~Aoki, D.~N. Sharma, and R.~Sankaranarayanan, ``Cancer of the
  cervix uteri,'' \emph{International journal of gynecology \& obstetrics},
  vol. 143, pp. 22--36, 2018.

\bibitem{Ronneberger-2015-UNet}
O.~Ronneberger, P.~Fischer, and T.~Brox, ``U-net: Convolutional networks for
  biomedical image segmentation,'' in \emph{Medical Image Computing and
  Computer-Assisted Intervention -- MICCAI 2015}, N.~Navab, J.~Hornegger, W.~M.
  Wells, and A.~F. Frangi, Eds., 2015, pp. 234--241.

\bibitem{Long-2015-FCN}
J.~Long, E.~Shelhamer, and T.~Darrell, ``Fully convolutional networks for
  semantic segmentation,'' in \emph{Proceedings of the IEEE conference on
  computer vision and pattern recognition}, 2015, pp. 3431--3440.

\bibitem{Zhou-2018-UNet++}
Z.~Zhou, M.~M. Rahman~Siddiquee, N.~Tajbakhsh, and J.~Liang, ``Unet++: A nested
  u-net architecture for medical image segmentation,'' in \emph{Deep Learning
  in Medical Image Analysis and Multimodal Learning for Clinical Decision
  Support}, D.~Stoyanov, Z.~Taylor, G.~Carneiro, T.~Syeda-Mahmood, A.~Martel,
  L.~Maier-Hein, J.~M.~R. Tavares, A.~Bradley, J.~P. Papa, V.~Belagiannis,
  J.~C. Nascimento, Z.~Lu, S.~Conjeti, M.~Moradi, H.~Greenspan, and
  A.~Madabhushi, Eds., 2018, pp. 3--11.

\bibitem{Badrinarayanan-2017-SegNet}
V.~Badrinarayanan, A.~Kendall, and R.~Cipolla, ``Segnet: A deep convolutional
  encoder-decoder architecture for image segmentation,'' \emph{IEEE
  transactions on pattern analysis and machine intelligence}, vol.~39, no.~12,
  pp. 2481--2495, 2017.

\bibitem{Chen-2021-TransUNet}
J.~Chen, Y.~Lu, Q.~Yu, X.~Luo, E.~Adeli, Y.~Wang, L.~Lu, A.~L. Yuille, and
  Y.~Zhou, ``Transunet: Transformers make strong encoders for medical image
  segmentation,'' \emph{CoRR}, 2021.

\bibitem{Vaswani-2017-Transformer}
A.~Vaswani, N.~Shazeer, N.~Parmar, J.~Uszkoreit, L.~Jones, A.~N. Gomez,
  {\L}.~Kaiser, and I.~Polosukhin, ``Attention is all you need,''
  \emph{Advances in neural information processing systems}, vol.~30, 2017.

\bibitem{Cao-2023-Swin-Unet}
H.~Cao, Y.~Wang, J.~Chen, D.~Jiang, X.~Zhang, Q.~Tian, and M.~Wang,
  ``Swin-unet: Unet-like pure transformer for medical image segmentation,'' in
  \emph{Computer Vision -- ECCV 2022 Workshops}, L.~Karlinsky, T.~Michaeli, and
  K.~Nishino, Eds.\hskip 1em plus 0.5em minus 0.4em\relax Cham: Springer Nature
  Switzerland, 2023, pp. 205--218.

\bibitem{Redmon-2018-YOLOv3}
J.~Redmon and A.~Farhadi, ``Yolov3: An incremental improvement,'' \emph{CoRR},
  vol. abs/1804.02767, 2018.

\bibitem{Bochkovskiy-2020-YOLOv4}
A.~Bochkovskiy, C.~Wang, and H.~M. Liao, ``Yolov4: Optimal speed and accuracy
  of object detection,'' \emph{CoRR}, vol. abs/2004.10934, 2020.

\bibitem{Zhu-2021-YOLOv5}
X.~Zhu, S.~Lyu, X.~Wang, and Q.~Zhao, ``Tph-yolov5: Improved yolov5 based on
  transformer prediction head for object detection on drone-captured
  scenarios,'' \emph{CoRR}, vol. abs/2108.11539, 2021.

\bibitem{Liu-2016-SSD}
W.~Liu, D.~Anguelov, D.~Erhan, C.~Szegedy, S.~Reed, C.-Y. Fu, and A.~C. Berg,
  ``Ssd: Single shot multibox detector,'' in \emph{Computer Vision -- ECCV
  2016}, B.~Leibe, J.~Matas, N.~Sebe, and M.~Welling, Eds.\hskip 1em plus 0.5em
  minus 0.4em\relax Cham: Springer International Publishing, 2016, pp. 21--37.

\bibitem{Ren-2015-FasterRCNN}
S.~Ren, K.~He, R.~Girshick, and J.~Sun, ``Faster r-cnn: Towards real-time
  object detection with region proposal networks,'' \emph{Advances in neural
  information processing systems}, vol.~28, 2015.

\bibitem{Lin-2017-RetinaNet}
\BIBentryALTinterwordspacing
T.~Lin, P.~Goyal, R.~B. Girshick, K.~He, and P.~Doll{\'{a}}r, ``Focal loss for
  dense object detection,'' \emph{CoRR}, vol. abs/1708.02002, 2017. [Online].
  Available: \url{http://arxiv.org/abs/1708.02002}
\BIBentrySTDinterwordspacing

\bibitem{Girshick-2014-RCNN}
R.~Girshick, J.~Donahue, T.~Darrell, and J.~Malik, ``Rich feature hierarchies
  for accurate object detection and semantic segmentation,'' in
  \emph{Proceedings of the IEEE conference on computer vision and pattern
  recognition}, 2014, pp. 580--587.

\bibitem{Girshick-2015-FastRCNN}
R.~Girshick, ``Fast r-cnn,'' in \emph{Proceedings of the IEEE international
  conference on computer vision}, 2015, pp. 1440--1448.

\bibitem{Dosovitskiy-2020-ViT}
\BIBentryALTinterwordspacing
A.~Dosovitskiy, L.~Beyer, A.~Kolesnikov, D.~Weissenborn, X.~Zhai,
  T.~Unterthiner, M.~Dehghani, M.~Minderer, G.~Heigold, S.~Gelly, J.~Uszkoreit,
  and N.~Houlsby, ``An image is worth 16x16 words: Transformers for image
  recognition at scale,'' \emph{CoRR}, vol. abs/2010.11929, 2020. [Online].
  Available: \url{https://arxiv.org/abs/2010.11929}
\BIBentrySTDinterwordspacing

\bibitem{Lin-2017-FPN}
T.-Y. Lin, P.~Doll{\'a}r, R.~Girshick, K.~He, B.~Hariharan, and S.~Belongie,
  ``Feature pyramid networks for object detection,'' in \emph{Proceedings of
  the IEEE conference on computer vision and pattern recognition}, 2017, pp.
  2117--2125.

\end{thebibliography}

\end{document}